\documentclass{INTERSPEECH2023}

\usepackage{booktabs} %
\usepackage{tabularx}
\usepackage{multirow} %
\usepackage{diagbox} %
\usepackage{color}
\usepackage[table]{xcolor}%
\usepackage{enumerate}
\usepackage[T1]{fontenc}
\usepackage{algpseudocode}%
\usepackage{algorithm}%
\usepackage{setspace}
\usepackage{threeparttable}
\DeclareMathOperator{\cnn}{CNN-Encoder}

\usepackage{tikz}
\newcommand{\tikzxmark}{%
\tikz[scale=0.23] {
    \draw[line width=0.7,line cap=round] (0,0) to [bend left=6] (1,1);
    \draw[line width=0.7,line cap=round] (0.2,0.95) to [bend right=3] (0.8,0.05);
}}
\newcommand{\tikzcmark}{%
\tikz[scale=0.23] {
    \draw[line width=0.7,line cap=round] (0.25,0) to [bend left=10] (1,1);
    \draw[line width=0.8,line cap=round] (0,0.35) to [bend right=1] (0.23,0);
}}

\renewenvironment{thebibliography}[1]{
  \begin{oldthebibliography}{#1}
  \setlength{\itemsep}{0.em}
  \setlength{\parskip}{0.em}
}
{
  \end{oldthebibliography}
}

\makeatletter
\def\bstctlcite{\@ifnextchar[{\@bstctlcite}{\@bstctlcite[@auxout]}}
\def\@bstctlcite[#1]#2{\@bsphack
  \@for\@citeb:=#2\do{%
    \edef\@citeb{\expandafter\@firstofone\@citeb}%
    \if@filesw\immediate\write\csname #1\endcsname{\string\citation{\@citeb}}\fi}%
  \@esphack}
\makeatother

\interspeechcameraready

\title{Weakly-Supervised Speech Pre-training: A Case Study on Target Speech Recognition}
\name{Wangyou Zhang, Yanmin Qian$^{\dagger}$\thanks{$^{\dagger}$Corresponding author.}}
\address{
  MoE Key Lab of Artificial Intelligence, AI Institute\\
  X-LANCE Lab, Department of Computer Science and Engineering\\
  Shanghai Jiao Tong University, China}
\email{wyz-97@sjtu.edu.cn, yanminqian@sjtu.edu.cn}

\begin{document}
\bstctlcite{IEEEexample:BSTcontrol} %

\maketitle
 
\begin{abstract}
Self-supervised learning (SSL) based speech pre-training has attracted much attention for its capability of extracting rich representations learned from massive unlabeled data.
On the other hand, the use of weakly-supervised data is less explored for speech pre-training.
To fill this gap, we propose a weakly-supervised speech pre-training method based on speaker-aware speech data.
It adopts a similar training procedure to the widely-used masked speech prediction based SSL framework, while incorporating additional target-speaker enrollment information as an auxiliary input.
In this way, the learned representation is steered towards the target speaker even in the presence of highly overlapping interference, allowing potential applications to tasks such as target speech recognition.
Our experiments on Libri2Mix and WSJ0-2mix datasets show that the proposed model achieves significantly better ASR performance compared to WavLM, the state-of-the-art SSL model with denoising capability.
\end{abstract}

\vspace{-4pt}
\section{Introduction}
\label{sec:intro}

Recently, self-supervised learning (SSL) based pre-training has greatly advanced research progress in speech processing, showing great potential in a wide range of downstream speech tasks~\cite{Towards-Shor2020,SUPERB-Yang2021}.
Existing SSL models can be roughly grouped into three categories based on their pre-training objectives, i.e., generative~\cite{Unsupervised-Chung2019,Multi_task-Ravanelli2020,TERA-Liu2021,Non_Autoregressive-Liu2021}, contrastive~\cite{Representation-Oord2018,Wav2vec-Schneider2019,Wav2vec2_0-Baevski2020,Wav2vec_Switch-Wang2022}, and predictive~\cite{HuBERT-Hsu2021,WavLM-Chen2022,Data2vec-Baevski2022,MT4SSL-Ma2023} approaches.
These SSL models are usually pre-trained on massive \emph{unlabeled} data in an application-agnostic manner~\cite{Self_Supervised-Mohamed2022}, and then fine-tuned on downstream speech tasks by updating either the entire network or only a small amount of parameters~\cite{Exploring-Chen2022,Adapter-Wang2022}.
The learned representations are found to be versatile for a series of speech tasks such as automatic speech recognition (ASR)~\cite{Exploration-Chang2021}, text-to-speech (TTS)~\cite{VQTTS-Du2022}, speaker verification (SV)~\cite{Why-Chen2022}, speech enhancement (SE)~\cite{Investigating-Huang2022}, and so on.

While existing SSL models are highly effective in extracting rich representations from single-speaker utterances, the capability of eliminating the interference from overlapped speech is still limited~\cite{WavLM-Chen2022,Investigating-Huang2022}.
However, this capability is especially important when tackling the well-known cocktail party problem~\cite{Some-Cherry1953,Past-Qian2018}, where multiple talkers speak simultaneously in a noisy environment.
It is therefore natural to ask whether we can improve the current speech pre-training paradigm to take the above problem into consideration.
Recently, there are a few studies working on this direction. Chen \emph{et al.}~\cite{WavLM-Chen2022} proposes to augment the input speech by overlapping with background speech or noise to force the SSL model to learn masked speech denoising and prediction at the same time.
Wang \emph{et al.}~\cite{Adapter-Wang2022} proposes to explicitly predict multiple labels corresponding to all utterances in the input overlapped speech, enabling the SSL model to learn denoising and separation in parallel with masked speech prediction.
These efforts still fall within the self-supervised speech pre-training framework, where no additional information other than the speech itself is exploited.

In this paper, we aim to explore a new speech pre-training direction, namely weakly-supervised speech pre-training, which allows the use of additional information (weak labels) about the data to facilitate better speech pre-training.
While similar ideas have been explored in computer vision and natural language processing~\cite{Large_scale-Ghadiyaram2019,Pretrained-Xiong2020}, it has not been well studied in the area of speech processing.
Our motivation is that the collected speech data for pre-training may be annotated with meta information such as the relative speaker identity. %
While SSL methods generally omit such information during pre-training, it can be helpful for the related downstream tasks if we take advantage of it properly during pre-training.

Therefore, in this paper, we make a first attempt at weakly-supervised speech pre-training using speaker-aware data.
The only assumption is that all speech samples can be divided into $S$ groups, each corresponding to a different speaker.
The speaker division can be obtained from either clustering~\cite{Self_Supervised-Han2022} or the data annotation\footnote{In this paper, we adopt the data annotation to simplify the discussion, and leave the former for future investigation.}.
Based on this assumption, we present a novel speech pre-training model---Target-Speaker HuBERT, or TS-HuBERT for short.
It adopts the same masked speech prediction objective as proposed in HuBERT~\cite{HuBERT-Hsu2021}, a widely-used SSL model.
During pre-training, the main input speech is randomly mixed with speech from a different speaker, and an auxiliary speech sample from the same main speaker is provided to guide the modeling of the main speaker's speech.
Taking both waveforms as input, the proposed pre-training model learns to predict the discrete targets of masked frames while eliminating the interference in the overlapped speech.
Furthermore, we investigate the effectiveness of the proposed TS-HuBERT model in the downstream target speech recognition task, which aims to recognize the target speaker's speech in the overlapped speech.
Our experiments show that the proposed model achieves significantly better ASR performance compared to WavLM~\cite{WavLM-Chen2022}, the state-of-the-art SSL model with denoising capability.

\vspace{-5pt}
\section{Weakly-supervised speech pre-training}
\subsection{Model design}
\label{ssec:design}

\begin{figure}[t]
  \centering
  \includegraphics[width=\columnwidth]{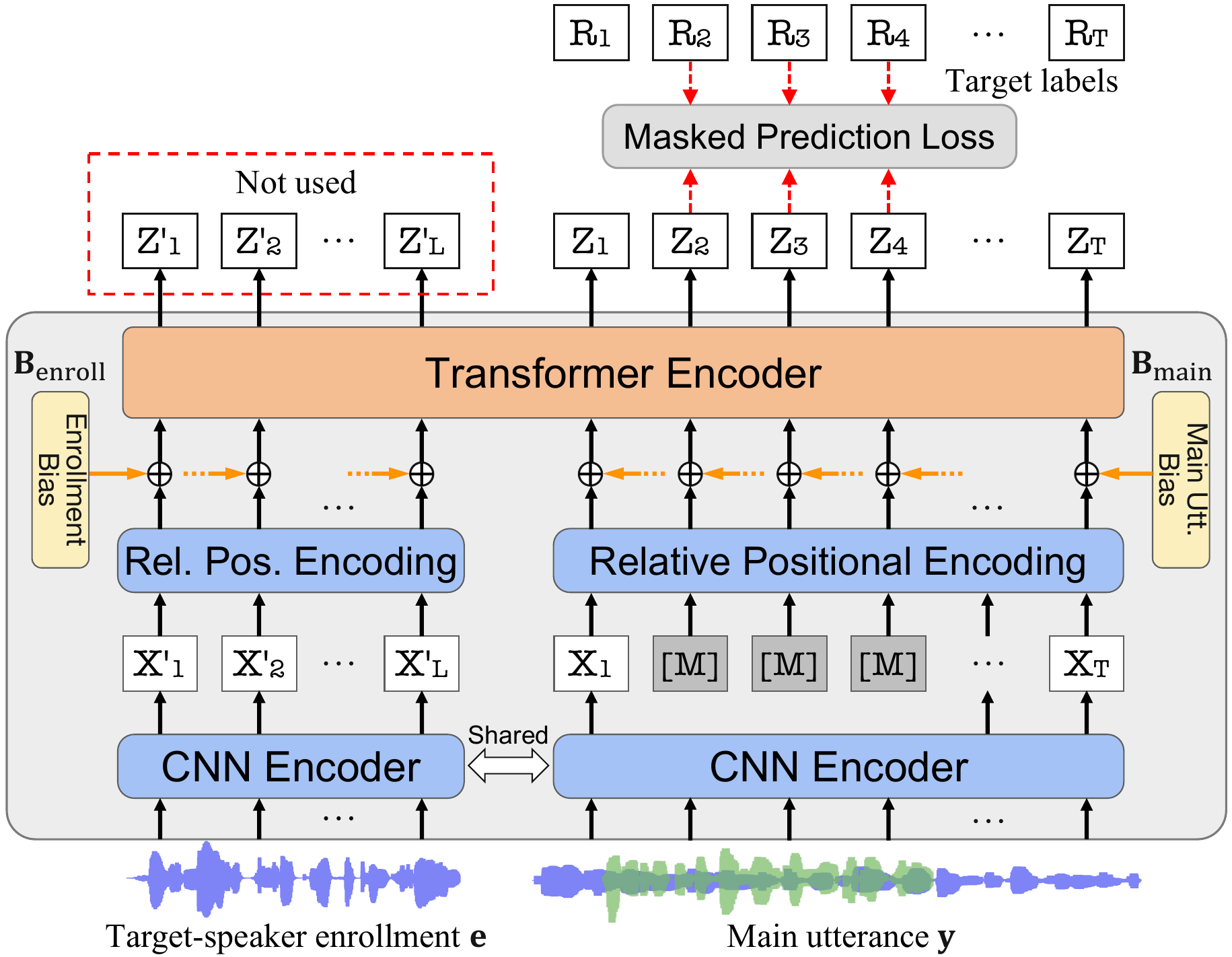}
  \captionsetup{labelfont=bf}
  \caption[ts_HuBERT]{Overview of the proposed weakly-supervised speech pre-training model based on speaker-aware speech data.}
  \label{fig:tsHuBERT}
  \vspace{-2pt}
\end{figure}

Figure~\ref{fig:tsHuBERT} presents the overview of the proposed TS-HuBERT model.
The entire model is built upon the well-established architecture in~\cite{Wav2vec2_0-Baevski2020,HuBERT-Hsu2021}, which consists of a convolutional neural network (CNN) encoder followed by a Transformer encoder.
In addition, following WavLM~\cite{WavLM-Chen2022}, the gated relative position bias~\cite{XLM_E-Chi2022} is also employed in the self-attention mechanism in Transformer encoder layers to boost the performance.
The input to the proposed model consists of a main utterance $\mathbf{y}$ for masked speech prediction~\cite{HuBERT-Hsu2021} and an auxiliary input $\mathbf{e}$ containing information about the target speaker.
The details of the pre-training data are presented in Section~\ref{ssec:data}.

Now, we consider integrating the target-speaker information with the aforementioned backbone.
In the literature, two types of data are often used to provide such information, i.e., an enrollment audio from the target speaker~\cite{End_to_End-Delcroix2019} and a speaker embedding vector~\cite{VoiceFilter-Wang2019}.
While the latter can directly provide a compact and efficient target-speaker representation, it entangles the pre-training model with a specific external speaker embedding model.
This inevitably reduces the generalizability and flexibility of the speech pre-training model.
Therefore, we opt for the raw enrollment audio as the auxiliary input and rely on the pre-training model to extract the target-speaker information implicitly.
Inspired by the success of temporal concatenation-based cross-modal modeling~\cite{InterBERT-Lin2020,Leveraging-Pan2022,Reading-Rahimi2022,icode-Wang2022}, we propose to fuse the two input streams via temporal concatenation.
More specifically, we first extract features for the enrollment $\mathbf{e}$ and the main utterance $\mathbf{y}$ using the same CNN encoder:
{\setlength\abovedisplayskip{3pt plus 0pt minus 7pt}
\setlength\belowdisplayskip{3pt plus 0pt minus 7pt}
\begin{align}
    \mathbf{X} &= \cnn(\mathbf{y}) &\in \mathbb{R}^{T \times D} \,, \label{eq:main_feat} \\
    \mathbf{X}^{\prime} &= \cnn(\mathbf{e}) &\in \mathbb{R}^{L \times D} \,, \label{eq:aux_feat}
\end{align}
}where $\mathbf{X}$ and $\mathbf{X}^{\prime}$ are the main utterance feature and the enrollment feature, respectively.
$T$ and $L$ are the corresponding numbers of feature frames, and $D$ is the feature dimension.
Later, we will concatenate the two features along the temporal dimension and leverage the Transformer encoder to learn the implicit correlation between them.
However, a simple concatenation of the two streams makes it difficult to distinguish them in the self-attention mechanism.
To mitigate this issue, we apply two independent convolution-based relative position encoding (rPE) layers~\cite{Transformers-Mohamed2019,Wav2vec2_0-Baevski2020} to inject the temporal order information into both streams.
In addition, we further enlarge the difference between the two sequences by adding a learnable bias vector to each of them, which is akin to the modality encoding in~\cite{Reading-Rahimi2022}:
{\allowdisplaybreaks
\setlength\abovedisplayskip{3pt plus 0pt minus 7pt}
\setlength\belowdisplayskip{3pt plus 0pt minus 7pt}
\begin{align}
    \mathbf{X}_{\text{in}} &= \operatorname{Conv}_{\text{rel\_pos}}(\mathbf{X}) + \mathbf{B}_{\text{main}} \,, \\
    \mathbf{X}^{\prime}_{\text{in}} &= \operatorname{Conv}^{\prime}_{\text{rel\_pos}}(\mathbf{X}^{\prime}) + \mathbf{B}_{\text{enroll}} \,,
\end{align}
}where $\mathbf{B}_{\text{main}} \in \mathbb{R}^{D}$ and $\mathbf{B}_{\text{enroll}} \in \mathbb{R}^{D}$ denote the main utterance bias vector and the enrollment bias vector, respectively.
The updated features $\mathbf{X}_{\text{in}}$ and $\mathbf{X}^{\prime}_{\text{in}}$ are finally concatenated and fed into the Transformer encoder to predict the frame-wise labels corresponding to the main utterance.
The output labels corresponding to the enrollment frames are discarded via slicing.

\vspace{-6pt}
\subsection{Speaker-aware masked speech prediction}
\label{ssec:objective}

In this subsection, we introduce the training objective of the proposed pre-training method, which follows the design in the HuBERT model~\cite{HuBERT-Hsu2021}.
The basic idea of the masked speech prediction objective in HuBERT is to predict the frame-wise discrete labels of the input speech while masking a portion of the feature frames generated by the CNN encoder.
The prediction loss (cross-entropy loss) is only calculated for the masked frames, while the labels are generated by one or more iterations of K-means clustering on the feature of the input speech.
We refer to the original paper~\cite{HuBERT-Hsu2021} for the details about label generation.

Since our major target is to improve the interference elimination ability of the speech pre-training model via weak supervision, we extend the original objective in HuBERT to fit this goal.
Given a main utterance input $\mathbf{y}$ and a target-speaker enrollment $\mathbf{e}$, we only apply masking to the main utterance feature $\mathbf{X}$ obtained in Eq.~(\ref{eq:main_feat}) and leave the enrollment feature $\mathbf{X}^{\prime}$ intact.
During pre-training, we also calculate the cross entropy loss only for the main utterance input, whose label is prepared in advance as in HuBERT training.
This is to encourage the model to focus on target speaker extraction with the full enrollment information.
Furthermore, to take the cocktail party problem into account, we augment the main utterance $\mathbf{y}$ by mixing it with a randomly sampled utterance from another speaker.
The detailed procedure is described in Section~\ref{ssec:data}.

In our implementation, the masks have a fixed length of 10 frames, and the number of masks is proportional to the main utterance length.
The maximum percentage of masked frames is 80\% and is usually not reached due to mask overlaps.
The masked feature frames are simply replaced with zeros.

\vspace{-6pt}
\subsection{Pre-training data preparation}
\label{ssec:data}

Similar to WavLM~\cite{WavLM-Chen2022}, we propose a speaker-aware utterance mixing strategy to simulate the overlapped pre-training data on the fly.
The core algorithm is illustrated in Algorithm~\ref{alg:simulation}.
The main differences compared to WavLM are:
(1) We sample the interference speech from the entire database instead of the current batch.
(2) We set the range of the overlap ratio to $[0, 100\%]$ instead of $[0, 50\%]$.
(3) An additional utterance $\mathbf{e}$ from the same main speaker is added to provide the target-speaker information.

\begin{algorithm}[th]
\setstretch{0.91}
\caption{Speaker-aware utterance mixing strategy}
\label{alg:simulation}
\textbf{Input: } an inventory of speaker-aware speech grouped by \\
\hphantom{\textbf{Input: xxx}} speaker IDs: $\mathbf{G}$; \\
\hphantom{\textbf{Input: }} a list of speaker IDs for all speech samples: $\mathbf{Q}$; \\
\hphantom{\textbf{Input: }} a list of pre-training labels for all speech samples: $\mathbf{R}$; \\
\hphantom{\textbf{Input: }} batch size: $i$; \\
\textbf{Output: } a batch of pre-training data: $\mathbf{U}$.
\begin{algorithmic}[1]
    \State $\mathbf{U} = \{\}$ %
    \State Sample $i$ utterances $\mathbf{U}_{\text{main}}$ uniformly from $\mathbf{G}$
    \For {each main utterance $\mathbf{y} \in \mathbf{U}_{\text{main}}$}
        \State $q_{\text{main}}$ $\leftarrow \mathbf{Q}[\mathbf{y}]$ \Comment{\footnotesize\emph{Find the speaker ID of} $\mathbf{y}$\normalsize}
            \State Sample a speaker ID $q\,(\neq q_{\text{main}})$ from $\mathbf{Q}$
            \State Sample an utterance $\mathbf{u_{\text{interf}}}$ from $\mathbf{G}[q]$ \Comment{\footnotesize\emph{Utts of speaker} $q$\normalsize}
            \State Sample the mixing energy ratio $k \sim \mathcal{U}(-5, 5)$
            \State Rescale $\mathbf{u_{\text{interf}}}$ such that $10 \log_{10}\frac{\Vert\mathbf{u_{\text{main}}}\Vert^2}{\Vert\mathbf{u_{\text{interf}}}\Vert^2} = k$
            \State $M \leftarrow \operatorname{length}(\mathbf{y})$; $\quad N \leftarrow \operatorname{length}(\mathbf{u}_{\text{interf}})$
            \State Sample the mixing length $l$ from $\{1, 2, \cdots, M\}$
            \State $l \leftarrow \min(l, N)$
            \State Sample the start position $m$ from $\{0, 1, \cdots, M-l\}$
            \State Sample the start position $n$ from $\{0, 1, \cdots, N-l\}$
            \State $\mathbf{y}[m:m+l] \leftarrow \mathbf{y}[m:m+l] + \mathbf{u}_{\text{interf}}[n:n+l]$
        \State Sample an enrollment $\mathbf{e}\, (\neq \mathbf{y})$ from $\mathbf{G}[q_{\text{main}}]$ %
        \State $\mathbf{r} \leftarrow \mathbf{R}[\mathbf{y}]$ \Comment{\footnotesize\emph{Find the label of} $\mathbf{y}$\normalsize}
        \State $\mathbf{U} \leftarrow \mathbf{U} \cup \{(\mathbf{y}, \mathbf{e}, \mathbf{r})\}$ \Comment{\footnotesize\emph{Append a new sample}\normalsize}
    \EndFor\\
    \Return $\mathbf{U}$
\end{algorithmic}
\end{algorithm}

\vspace{-6pt}
\subsection{Application: target speech recognition}
\label{ssec:application}

In this subsection, we introduce several fine-tuning methods to apply the proposed TS-HuBERT model to the downstream target speech recognition task.
It aims to recognize the target speaker's speech in the overlapped speech, which is a typical task in the cocktail party problem.
One straightforward fine-tuning approach is using a linear projection layer on top of the pre-training model to map the feature to the output dimensionality (vocabulary size)~\cite{Wav2vec2_0-Baevski2020,HuBERT-Hsu2021}.
And the connectionist temporal classification (CTC) loss~\cite{Connectionist-Graves2006} is used for end-to-end training.
In this procedure, the entire pre-training model (except for the CNN encoder) will be updated to fit the downstream task.
Since TS-HuBERT can extract the target-speaker information from the enrollment, it naturally fits this fine-tuning method.

For SSL models that cannot utilize the enrollment directly, another adaptation-based fine-tuning approach~\cite{Adapting-Huang2022} can be used, where lightweight speaker adaptation layers are inserted into the pre-training model for joint fine-tuning.
The newly inserted layers take as input a pre-extracted speaker embedding vector $\mathbf{e}_{\text{emb}}$ to steer the intermediate representations in pre-training models towards the target speaker.
Here, we evaluate the following three adaptation layers proposed in ~\cite{Adapting-Huang2022} with TS-HuBERT:
\begin{itemize}
\setlength{\itemindent}{3pt}
    \item[1.] \textbf{Add:} $\mathbf{e}_{\text{emb}}$ is directly added to the CNN encoder output $\mathbf{X}$ through a linear projection to match the hidden dimension.
    \item[2.] \textbf{FiLM:} $\mathbf{e}_{\text{emb}}$ is used to estimate a feature-wise linear modulation (FiLM)~\cite{FiLM-Perez2018} for the CNN output $\mathbf{X}$:
{\setlength\abovedisplayskip{5pt plus 0pt minus 7pt}
\setlength\belowdisplayskip{5pt plus 0pt minus 7pt}
    \begin{align}
        \mathbf{X} \Leftarrow w(\mathbf{e}_{\text{emb}}) \cdot \mathbf{X} + b(\mathbf{e}_{\text{emb}}) \,,
    \end{align}
    }where $w(\cdot)$ and $b(\cdot)$ are two linear projection layers.
    \item[3.] \textbf{cLN:} $\mathbf{e}_{\text{emb}}$ is used to estimate FiLM transformations for the layer normalizations (LNs) in the first Transformer encoder layer, converting them into conditional LNs (cLNs)~\cite{Conditionally-Pilault2021}:
{\setlength\abovedisplayskip{3pt plus 0pt minus 7pt}
\setlength\belowdisplayskip{3pt plus 0pt minus 7pt}
    \begin{align}
        \mathbf{X}_{\text{cLN}} = \big[w(\mathbf{e}_{\text{emb}}) \cdot \boldsymbol{\gamma} + b(\mathbf{e}_{\text{emb}})\big] \cdot \dfrac{\mathbf{X} - \boldsymbol{\mu}}{\boldsymbol{\sigma}} + \boldsymbol{\beta} \,,
    \end{align}
    }where $\boldsymbol{\gamma}$ and $\boldsymbol{\beta}$ are the learnable scaling and biasing parameters in standard LNs, respectively.
    $\boldsymbol{\mu}$ and $\boldsymbol{\sigma}$ are the mean and standard deviation of the input $\mathbf{X}$, respectively.
\end{itemize}
We apply the near-identity initialization~\cite{Parameter_efficient-KarimiMahabadi2021} to all adaptation layers, which is found crucial in our preliminary experiments.

\vspace{-8pt}
\section{Experiments}

\vspace{-4pt}
\subsection{Experimental setup}
\label{ssec:setup}
\vspace{-4pt}

Our experiments were done using the fairseq toolkit\footnote{\texttt{github.com/facebookresearch/fairseq\,(313ff05)}}.
The TS-HuBERT model follows the same structure as the WavLM Base model, which enhances the HuBERT~\cite{HuBERT-Hsu2021} Transformer encoder by employing the gated relative position bias~\cite{XLM_E-Chi2022} in the self-attention mechanism.
Due to the space limitation, we refer to the WavLM paper~\cite{WavLM-Chen2022} for the detailed parameters.
As described in Section~\ref{ssec:design}, we additionally insert two convolution-based rPE layers and two 768-dimensional learnable bias vectors before the Transformer encoder.
Each rPE layer consists of a 16-group convolution layer with kernel size 128 and the Gaussian error linear unit (GELU) activation.
These new layers account for 10M new parameters, resulting in 104.37M parameters in total.
The TS-HuBERT model is pre-trained for one iteration (400k steps) on the 960-hour Librispeech data~\cite{Librispeech-Panayotov2015} using labels generated by 500-centroid K-means clustering of the 9-th Transformer layer output of the HuBERT Base model\footnote{Available at \url{https://dl.fbaipublicfiles.com/HuBERT/HuBERT_base_ls960.pt}}.
The batch size is at most 312 seconds of audio per GPU.
Other hyperparameters are the same as those in HuBERT Base~\cite{HuBERT-Hsu2021}.
The enrollment $\mathbf{e}$ is randomly truncated to 48000 samples to avoid massive memory consumption during pre-training.

As for fine-tuning on downstream tasks, we use the character-level CTC loss for all experiments.
We evaluate the performance of target speech recognition on two commonly-used datasets: noisy Libri2Mix~\cite{LibriMix-Cosentino2020} and WSJ0-2mix~\cite{Deep-Hershey2016}.
The sample rate of all speech data is 16 kHz.
In Libri2Mix, there are 13900, 3000, and 3000 samples in the training\footnote{We only use the \texttt{train-100} subset to speed up experiments.}, validation, and evaluation sets, respectively.
In WSJ0-2mix, the numbers of samples are 20000, 5000, and 3000, respectively.
The sets of speakers in training and evaluation sets do not overlap.
We adopt the existing enrollment lists for samples in the validation and evaluation sets of Libri2Mix\footnote{\url{https://github.com/BUTSpeechFIT/speakerbeam}} and WSJ0-2mix\footnote{\url{https://github.com/gemengtju/SpEx_Plus}}, respectively.
For the training sets, we randomly select the same-speaker enrollment for each sample to increase diversity.
The batch size for fine-tuning is 125 seconds of audio per GPU.
The peak learning rate is 2e-5 with 8000 warmup steps.
During fine-tuning, we always freeze the CNN encoder in the pre-training model.
For adaptation-based fine-tuning, we use a ResNet-34 model~\cite{Deep-He2016,Build-Chen2022} pre-trained on the VoxCeleb corpus~\cite{VoxCeleb-Nagrani2017} to extract 256-dimensional speaker embeddings $\mathbf{e}_{\text{emb}}$.
During evaluation, we adopt the Viterbi algorithm for decoding, and the word error rate (WER) is calculated as the metric.
Unless specifically mentioned, no language model is used.
We used 8 RTX 2080 Ti GPUs for all our experiments.\footnote{Code is available at \url{https://github.com/Emrys365/fairseq/tree/wavlm/examples/tshubert}.}

\vspace{-6pt}
\subsection{Evaluation on standard speech recognition}
\vspace{-2pt}
Before applying the proposed pre-training model to target speech recognition, we are firstly interested in the performance of the proposed model in the standard ASR task.
To this end, we fine-tune the TS-HuBERT model on three different subsets of Librispeech as in~\cite{Wav2vec2_0-Baevski2020,HuBERT-Hsu2021,WavLM-Chen2022} and evaluate it on the \texttt{test-clean} and \texttt{test-other} subsets.
We simply follow the fine-tuning configuration as in~\cite{HuBERT-Hsu2021,WavLM-Chen2022} and use a batch size of 200s of audio per GPU.
Since no enrollment is available in the standard ASR task, we discard the enrollment-related processing and only feed the main utterance feature into the Transformer encoder (denoted as ``w/o $\mathbf{e}$'').
We compare three speech pre-training models in Table~\ref{tab:exp_librispeech}, where all models are pre-trained on the 960-hour Librispeech data for 400k steps.
Interestingly, although our proposed model is pre-trained always with an additional enrollment input, it can still be fine-tuned to achieve comparable performance compared to the two SSL models.
This reveals the potential of the proposed TS-HuBERT model to adapt to different tasks.

\begin{table}[t]
    \caption{WER (\%) evaluation of speech pre-training  models for standard speech recognition on the Librispeech test sets. Note that no language model is used for the listed models.}
    \label{tab:exp_librispeech}
    \centering
    \begin{tabular}{lcc}
        \toprule[1pt]
        \textbf{Method} & \texttt{test-clean} & \texttt{test-other} \\
        \midrule
        \multicolumn{3}{c}{\textbf{\emph{1-hour labeled}}} \\
        wav2vec 2.0 Base~\cite{WavLM-Chen2022} & 24.5 & 29.7 \\
        WavLM Base~\cite{WavLM-Chen2022} & 24.5 & 29.2 \\
        TS-HuBERT (w/o $\mathbf{e}$) & \textbf{20.8} & \textbf{28.1} \\
        \midrule
        \multicolumn{3}{c}{\textbf{\emph{10-hour labeled}}} \\
        wav2vec 2.0 Base~\cite{WavLM-Chen2022} & 11.1 & 17.6 \\
        WavLM Base~\cite{WavLM-Chen2022} & \textbf{9.8} & \textbf{16.0} \\
        TS-HuBERT (w/o $\mathbf{e}$) & 10.7 & 18.9 \\
        \midrule
        \multicolumn{3}{c}{\textbf{\emph{100-hour labeled}}} \\
        wav2vec 2.0 Base~\cite{WavLM-Chen2022} & 6.1 & 13.3 \\
        WavLM Base~\cite{WavLM-Chen2022} & \textbf{5.7} & \textbf{12.0} \\
        TS-HuBERT (w/o $\mathbf{e}$) & 6.0 & 13.9 \\
        \bottomrule[1pt]
    \end{tabular}
\end{table}

\vspace{-6pt}
\subsection{Evaluation on target speech recognition}
\vspace{-2pt}
Next, we evaluate the performance of the speech pre-training models on the target speech recognition task.
We first conduct experiments on the WSJ0-2mix dataset.
The results are shown in Table~\ref{tab:exp_wsj0_2mix}.
We first compare with two state-of-the-art multi-speaker ASR methods on WSJ0-2mix, i.e., the jointly trained speech separation and ASR models~\cite{Train-Shi2022} (denoted as ``DPRNN-ASR''), and a monolithic multi-speaker ASR model~\cite{Exploration-Chang2021} (denoted as ``PIT-ASR'').
For the SSL-based method, we compare with the WavLM Base model as it is the state-of-the-art SSL model with denoising capability and has a similar pre-training setup to ours.
The same adaptation-based fine-tuning approaches as described in Section~\ref{ssec:application} are applied to the WavLM Base model to enable target speech recognition.

From Table~\ref{tab:exp_wsj0_2mix}, we can see that directly fine-tuning TS-HuBERT without adaptation layers can readily achieve very promising results (WER<7\%).
The adaptation-based fine-tuning methods also work well with both WavLM Base and TS-HuBERT, and increasing the number of fine-tuning steps can slightly improve the performance.
In~\cite{HuBERT-Hsu2021,WavLM-Chen2022}, the Transformer encoder in the pre-training model is often frozen for the first 10k fine-tuning steps.
Here, we empirically find this trick is harmful to the target speech recognition performance, and it is thus disabled for the rest experiments.
It is noteworthy that our proposed TS-HuBERT model obtains the new state-of-the-art performance (WER=6.1\%) on WSJ0-2mix, which is significantly better than the WavLM Base model.

Furthermore, we evaluate the models on the Libri2Mix dataset, which is much more difficult due to the presence of noise.\footnote{Differently,~\cite{Adapting-Huang2022} uses the clean version of Libri2Mix.}
Due to the space limitation, we only present the best results for the adaptation-based fine-tuning methods.
In addition, we compare with the monolithic multi-speaker ASR method (denoted as ``PIT-ASR'') from ESPnet~\cite{ESPnet-Watanabe2018}, and enhance it by adding more training data (denoted as ``+ \texttt{train-360}'') and by further applying speed perturbation~(SP)~\cite{Audio-Ko2015} and a Transformer language model (LM).
Although the pre-training-based models only use the \texttt{train-100} subset for fine-tuning, they obtain substantial performance improvement over the PIT-ASR approach, even when the latter uses much more training data (``+ \texttt{train-360}'').
The proposed TS-HuBERT again surpasses the WavLM Base model, with \textasciitilde 10\% relative WER reduction.
However, their performance still lags behind the doubly enhanced PIT-ASR approach (``++ SP \& LM''), implying room for further improvement in the noisy scenario.

\begin{table}[th]
    \caption{Evaluation of different speech pre-training  models for target speech recognition on the WSJ0-2mix test set. ``\#Param'' denotes the number of trainable parameters. ``Frz. 10k'' denotes whether or not to freeze the pre-training model for the first 10k steps. ``50k'' and ``100k'' denote the fine-tuning steps.}
    \label{tab:exp_wsj0_2mix}
    \centering
    \resizebox{\columnwidth}{!}{%
    \setstretch{0.8}
    \begin{tabular}{lccc|cc}
        \toprule[1pt]
        \multirow{2}{*}{\textbf{Method}} & \multirow{2}{*}{\textbf{\shortstack[c]{Adapt.\\Type}}} & \multirow{2}{*}{\textbf{\#Param}} & \multirow{2}{*}{\textbf{\shortstack[c]{Frz.\\10k}}} & \multicolumn{2}{c}{\textbf{WER\,(\%)}} \\
        & & & & 50k & 100k \\
        \midrule
         \multicolumn{2}{l}{DPRNN-ASR (8 kHz)~\cite{Train-Shi2022}} & - & - & \multicolumn{2}{c}{\textbf{7.1}} \\
         PIT-ASR~\cite{Exploration-Chang2021} & - & - & - & \multicolumn{2}{c}{12.1} \\
        \midrule
         \multirow{6}{*}{WavLM Base} & \multirow{2}{*}{Add} & \multirow{2}{*}{94.75 M} & $\tikzcmark$ & 16.6 & 14.6 \\
         &  &  & $\tikzxmark$ & 11.7 & 11.3 \\
         \cline{2-6}
         & \multirow{2}{*}{FiLM} & \multirow{2}{*}{94.95 M} & $\tikzcmark$ & 15.8 & 13.9 \\
         &  &  & $\tikzxmark$ & 11.7 & 11.2 \\
         \cline{2-6}
         & \multirow{2}{*}{cLN} & \multirow{2}{*}{95.34 M} & $\tikzcmark$ & 13.0 & 12.4 \\
         &  &  & $\tikzxmark$ & 10.9 & 10.4 \\
        \midrule
         \multirow{8}{*}{TS-HuBERT} & \multirow{2}{*}{-} & \multirow{2}{*}{104.39 M} & $\tikzcmark$ & 6.9 & 6.5 \\
         &  &  & $\tikzxmark$ & 6.3 & \textbf{6.1} \\
         \cline{2-6}
         & \multirow{2}{*}{Add} & \multirow{2}{*}{104.59 M} & $\tikzcmark$ & 7.4 & 6.8 \\
         &  &  & $\tikzxmark$ & \textbf{6.2} & 6.1 \\
         \cline{2-6}
         & \multirow{2}{*}{FiLM} & \multirow{2}{*}{104.78 M} & $\tikzcmark$ & 7.0 & 6.5 \\
         &  &  & $\tikzxmark$ & 6.3 & 6.1 \\
         \cline{2-6}
         & \multirow{2}{*}{cLN} & \multirow{2}{*}{105.18 M} & $\tikzcmark$ & 6.8 & 6.3 \\
         &  &  & $\tikzxmark$ & 6.4 & 6.1 \\
        \bottomrule[1pt]
    \end{tabular}%
    }
\end{table}

\begin{table}[th]
    \caption{Evaluation of different speech pre-training  models for target speech recognition on the noisy Libri2mix test set. ``\#Param'' denotes the number of trainable parameters. ``100k'' and ``250k'' denote the fine-tuning steps.}
    \label{tab:exp_libri2mix}
    \centering
    \setstretch{0.9}
    \begin{tabular}{lcc|cc}
        \toprule[1pt]
        \multirow{2}{*}{\textbf{Method}} & \multirow{2}{*}{\textbf{\shortstack[c]{Adapt.\\Type}}} & \multirow{2}{*}{\textbf{\#Param}} & \multicolumn{2}{c}{\textbf{WER\,(\%)}} \\
        & & & 100k & 250k \\
        \midrule
        PIT-ASR & - & 32 M & \multicolumn{2}{c}{50.1} \\
        \hspace{1em}+ \texttt{train-360} & - & 32 M & \multicolumn{2}{c}{31.0} \\
        \hspace{2em} ++ SP \& LM & - & 32 M & \multicolumn{2}{c}{\textbf{23.5}} \\
        \midrule
        WavLM Base & cLN & 95.34 M & 28.5 & 27.5 \\
        \midrule
         \multirow{2}{*}{TS-HuBERT}  & - & 104.39 M & 25.5 & 24.9 \\
         & cLN & 105.18 M & 25.5 & 24.8 \\
        \bottomrule[1pt]
    \end{tabular}%
\end{table}

\vspace{-4pt}
\section{Conclusions}
\vspace{-6pt}
In this paper, we propose a novel weakly-supervised speech pre-training model---TS-HuBERT, which utilizes an additional enrollment to incorporate the target-speaker information.
Pre-trained with the masked speech prediction objective on overlapped speech, TS-HuBERT learns to eliminate the interference by leveraging the enrollment information.
Experiments show that TS-HuBERT can work well on both standard ASR and target speech recognition tasks.
In future work, we would like to extend TS-HuBERT with clustering-based speaker labels for pre-training, thus making it better fit realistic applications.

\vspace{-8pt}
\section{Acknowledgements}
\vspace{-4pt}
The authors would like to thank Zili Huang from Johns Hopkins University for the helpful discussion on adapting SSL models to target speech recognition.
This work was supported in part by China STI 2030-Major Projects under Grant No. 2021ZD0201500, in part by China NSFC projects under Grants 62122050 and 62071288, and in part by Shanghai Municipal Science and Technology Major Project under Grant 2021SHZDZX0102.
Experiments have been carried out on the PI supercomputer at Shanghai Jiao Tong University.

\vspace{-8pt}
\bibliographystyle{IEEEtran}
\bibliography{mybib}

\end{document}